\documentclass[aps,prl,twocolumn,groupedaddress]{revtex4}

\usepackage{graphicx}

\newcommand{\wn}{${cm^{-1} }$}
\newcommand{\sigone}{$\sigma_{1}(\omega)$}
\newcommand{\sigo}{$\sigma(\omega)$}

\newcommand{\meV}{$meV$}

\newcommand{\K}{$K$}

\newcommand{\zwo}{ZrW${_2}$O${_8}$}
\newcommand{\hwo}{HfW${_2}$O${_8}$}
\newcommand{\ch}{CH${_3}$I}
\newcommand{\xch}{XCH${_3}$}
\newcommand{\wo}{WO${_4}$}

\newcommand{\degree}{^{\circ}}

\newcommand{\s}{\hspace{.05cm}}

\newcommand{\direction}[1]{$\langle #1\rangle$}

\begin{document}
\title{Unusual Low-Energy Phonon Dynamics in the Negative Thermal Expansion Compound \zwo}


\author{Jason N. \surname{Hancock}}
\author{Chandra \surname{Turpen}}
\author{Zack \surname{Schlesinger}}
\affiliation{Physics Department, University of California Santa Cruz, Santa Cruz, CA 95064, USA}
\author{Glen R. \surname{Kowach}}
\affiliation{Department of Chemistry, The City College of New York, New York, NY 10031}
\author{Arthur P. \surname{Ramirez}}
\affiliation{Bell Laboratories, Lucent Technologies, 600 Mountain Ave, Murray Hill, NJ 07974}


\date{\today}

\begin{abstract}
An infrared study of the phonon spectra of \zwo\  as a function of temperature
which includes the low energy (2-10\s\meV) region relevant to negative thermal expansion
is reported and discussed in the context of specific heat and neutron density of states results. 
The prevalence of infrared active phonons at low energy and their observed temperature 
dependence are highly unusual and indicative of exotic low-energy lattice dynamics.
Eigenvector calculations indicate a mixing of librational and translational motion within each low-frequency IR mode.
The role of the underconstrained structure in establishing the nature of these modes,
and the relationship between the IR spectra and the
large negative thermal expansion in \zwo\ are discussed.

\end{abstract}


\maketitle

The tendency of solids to expand when heated is one of the most pervasive and familiar
phenomena in solid state physics\cite{ashcroft},
however, there are some compounds in which the opposite phenomenon occurs\cite{white93,sle98}.
In \zwo\  this negative thermal expansion (NTE)  is large, isotropic and persists over a broad temperature
range\cite{mary96,evans96,pryde96,ram98,ern98,david99}.  
These characteristics enhance our interest in both the fundamental physics
of \zwo , and the potential applications of such a substantial NTE effect, which are widespread\cite{sle98}.
A defining feature of \zwo\ is the existence of a 1-fold coordinated oxygen site for each
\wo\ tetrahedron. This terminal, or unconstrained\cite{ern98}, oxygen creates a structural openness along
the high-symmetry \direction{111} axes and is thought to be influential in the low-energy
dynamics and crucial to the phenomenon of NTE in \zwo .
     
Thermal expansion phenomena generally come from
anharmonic phonon dynamics\cite{ashcroft,pryde96,ram98,ern98,david99,mit99}. 
For \zwo\ analysis of temperature dependence\cite{ram98,ern98,david99,mit99} indicates
that the low energy region should be of primary importance.
To probe the low-energy phonon dynamics we use temperature dependent infrared spectroscopy 
covering the range from 5000\s\wn\ to 16\s\wn\ (600-2\s\meV), which encompasses all the optic phonons of \zwo.
We find that the phonon-related peaks in the infrared spectra extend to unusually low energy ($\sim$ 3\s\meV),
and that there is strong and unusual temperature 
dependence in the low-energy region, reflecting evidence of unconventional and anharmonic behavior.
We suggest that understanding the mechanism of NTE and understanding the exotic nature of the infrared
spectra from \zwo\ are two essentially similar problems, both rooted in exotic low energy dynamics.  

One of the central unresolved issues for \zwo\ has to do with the nature of the eigenmode(s) responsible for
NTE.  While earlier work has emphasized the role of transverse 
oxygen vibration\cite{mary96,evans96,pryde96,ram98,ern98,david99}
for the mechanism of NTE, Cao et al.\cite{cao02,cao03} have proposed a model, based on XAFS data, 
in which translation of \wo\ tetrahedra along \direction{111} directions plays an important role.
Using eigenvector calculations together with our data, we infer that transverse oxygen motion,
in the form of libration, tends to be accompanied by translation 
within each low-energy mode (thus they are essentially inseparable).
This mixing arises due to the unconstrained oxygen and the associated openness of the structure 
along the high-symmmetry \direction{111} direction axes, and it appears to be essential to the nature of the 
low-frequency IR spectra.
This mixing may also be crucial to the mechanism of the NTE:
while the transverse oxygen motion provides the thermal motion that contracts
the lattice, the presence of a translational component in each low-energy mode provides a source of
frustration that inhibits lattice instabilities that would destroy NTE.

The single-crystal samples are grown using a layered self-flux technique\cite{kow00}.
Our measurements of the specular reflectivity cover the frequency range  from 15 to 5000$\s cm^{-1}$  and 
the temperature range 20 to $300\s K$. 
The optical conductivity, \sigo , is determined from the reflectivity via a Kramers-Kronig transform with carefully chosen upper and lower terminations. For the acentric $P2_13$ structure of \zwo\ one expects a total of 132 modes: 3 acoustic modes, 32 triplets of IR optical phonons, 11 doublets and 11 singlets, associated with the irreducible representations
$33T+11E+11A$\cite{rav00,rav01,chen01}.   
Our results are broadly consistent with earlier work\cite{rav00,rav01,chen01,yam02,rav03}  for the higher frequency
parts of the phonon spectrum, and complementary in  
our emphasis on the low frequency range.

\begin{figure}
\begin{center}
\includegraphics[width=3.4in]{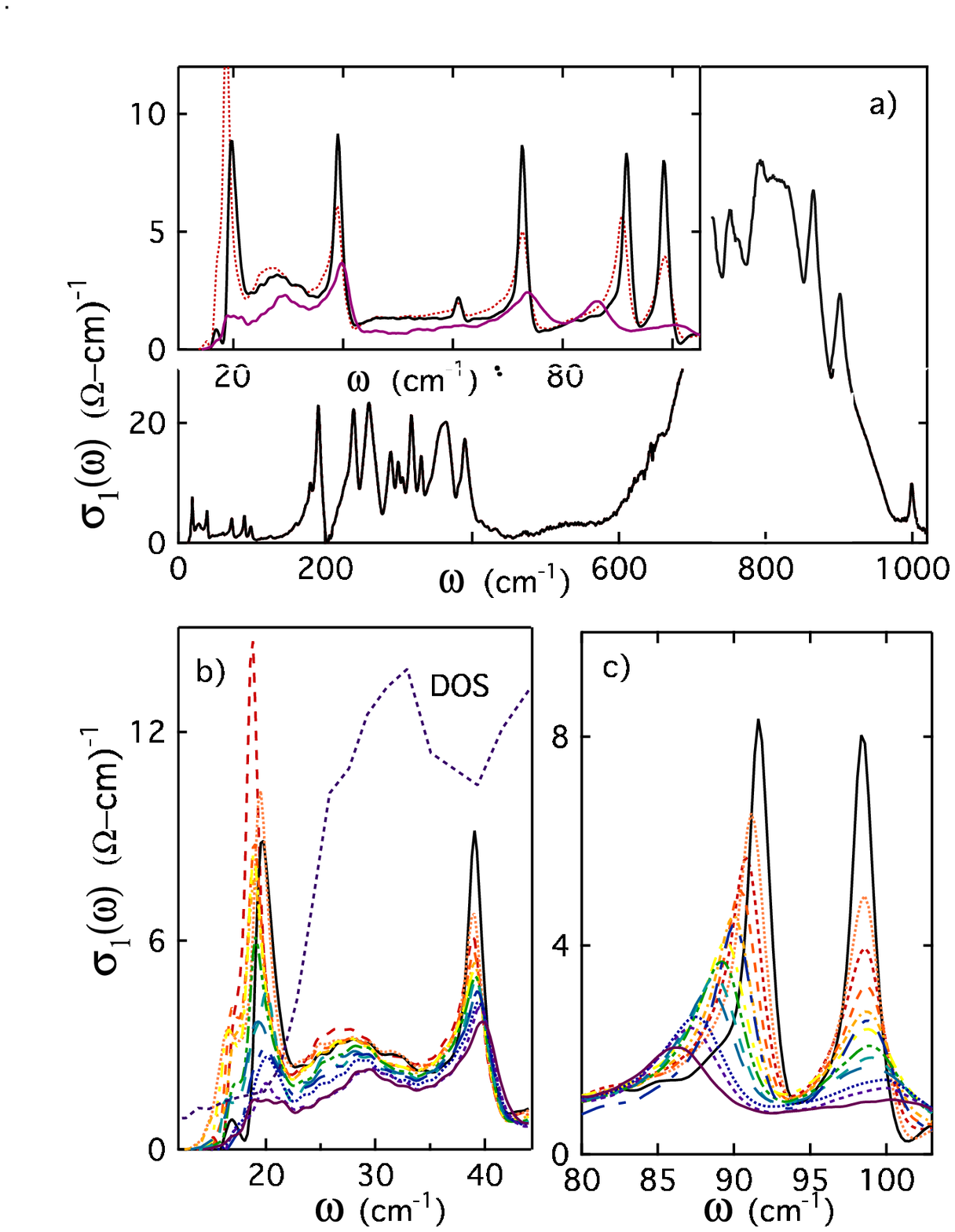}
\caption{(color online) a) The real part of the conductivity, \sigone , is shown for the entire optic phonon region.
The inset shows the low frequency region for temperatures of 20, 80(dotted) and 240 K.
b) and c) show \sigone\  for T=20, 60, 80, 100, 120, 140, 160, 180, 200, 220, 240, 260 and 300 K in two
low $\omega$ regions, along with neutron density of states in (b)\cite{ern98} }
\label{fig:ss}
\end{center}
\end{figure}

Figure 1a shows the real part of the optical conductivity, \sigone , vs frequency. The infrared active optical phonon modes, which generally appear as peaks in  \sigone , are sufficiently dense that they tend to merge together into clusters in \zwo . The upper cluster is associated with bond stretching motion (including the $\nu_3$ modes of the \wo\ tetrahedra) and the middle cluster (180-400\s\wn) with bond bending modes (e.g., $\nu_2$ modes).  
Below these two clusters
there are additional infrared features, shown in the inset,
which we associate with librational and translational motion of 
\wo\ tetrahedra, as discussed below.
Unlike the generic modes of the upper clusters, the modes in this region
are specific to the \zwo\ structure, and it is in this lower region that one expects
to find modes relevant to the mechanism of NTE. 

As shown in Fig. 1b) and c), the peaks in this region extend to unusually low energy and exhibit striking temperature dependence.
They tend to sharpen and grow in strength as the sample is cooled; they exhibit swept,
asymmetric lineshapes, and in some cases substantial dependence of peak frequency on temperature.  
The largest peak frequency shifts are observed for the broad 28\s\wn\ mode, which shifts from about 
27 to 29.5\s\wn\ (+8\%), and the 88\s\wn\ mode, which shifts from about 91.5 to 86\s\wn\ (-6\%). 
These shifts are unusually large and the substantial accretion of spectral weight 
(presumably coming from the higher frequency) 
is beyond the scope of conventional phonon models.
The pervasiveness of strong temperature dependence throughout this low frequency region reflects 
the exotic character of the low energy dynamics which is clearly evident in the IR spectra.

It is informative to consider the relationship between our IR data and specific heat\cite{ram98}.
For the fit shown in Figure 2a, a lowest energy Einstein mode
at 28.5\s\wn\ with a spectral weight of 2.6 oscillators per unit cell
provides a good fit to the leading edge of $C(T)/T^3$
\footnote{The fit includes a $\Theta _D =62$\s\wn\  and Einstein contributions at IR and Raman peak frequencies
above 28\s\wn .}.
This fit is highly constraining with regard to the energy of the lowest Einstein mode;
attempts to fit $C(T)$ with a lower optical mode
generate excess specific heat at low temperature as shown in Figure 2a. 
We therefore conclude that the 20\s\wn\ peak does not correspond
to the zone-center endpoint of an ordinary optical phonon branch and that the broad
28\s\wn\ peak is the signature of the lowest optical phonon of \zwo . The total spectral weight in the
peaks between 
28 and 100\s\wn\ is in agreement with the results of our phonon eigenvector calculations, discussed below,
and the peaks in this region are in reasonable correspondence with zone-center intercepts of 
optical phonon dispersion relations calculated for the isostructural compound \hwo \cite{mit01,mit03}
as well as neutron DOS peaks\cite{ern98}.
The 20\s\wn\ peak differs from the others in that it diminishes substantially below 80\s\K (the others grow)
and is not reconcileable with neutron DOS and $C(T)$.
These two features may suggest a low-density defect with
an extremely large dipole matrix element.  Further theory is needed, in conjunction
with controlled defect studies, to ascertain the origin of this marked absorption, and
its possible connection with the unusual lattice dynamics of the bulk compound.

\begin{figure}
\includegraphics[width=3.6in]{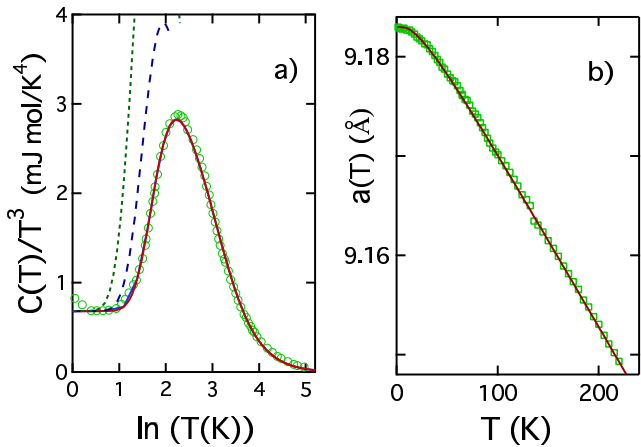}
\caption{(color online) a) A fit to the specific heat with a lowest infrared mode at 28.5\s\wn\  (3.5 meV) is shown by the solid lie (through the 
specific heat data).  The dashed curves show the excess specific heat generated by failed attempts to fit $C(T)$ with a lowest optic mode at 20 or 24\s\wn , respectively.
b) a(T) data (open squares) is shown along with a fit based on a two Einstein mode model.}
\label{fig:ss}
\end{figure}

Focusing on energetics, one can let the infrared spectra provide a context in which to explore the origins of the temperature dependence of the lattice parameter, a($T$). 
Similar to earlier work\cite{ern98,david99}, we approach this by associating negative Gr\"uneisen parameters ($\gamma$)
with Einstein mode frequencies.  
We calculate a($T$) using a bulk modulus of 72 GPa \cite{jorg99} and assuming triplet weight\cite{ashcroft}.   
Starting with the frequency of the lowest optic mode from our data (at 28\s\wn=3.5\s\meV), we find that associating a large negative Gr\"uneisen parameter to this mode alone does not allow a good fit to a($T$), however, a good fit can be obtained simply by including a second Einstein mode at higher frequency, as shown in Figure 2b) for  $\gamma$=-20 at 28\s\wn\ and $\gamma$=-7 at  88 \s\wn\ . The lower energy corresponds to the lowest energy optical phonon in \zwo , 
which is so low it cuts through the acoustic phonon dispersion and may thereby acquire mixed character\cite{mit03};
the higher energy corresponds to the strongly temperature dependent peak near 88 \s\wn\ (Fig 1c).
While this 2-mode approach provides the minimal context in which a good fit can be obtained,
one can also do more nuanced fits involving additional modes.
One can use Gr\"uneisen parameters for modes above 40\s\wn\
from pressure dependent Raman measurements of Ravindran et al.\cite{rav01} 
and add an additional term for the 28 \s\wn\ mode.   In that more nuanced approach
good fits to a($T$) can be obtained when the $\gamma$'s of the two lowest optic modes add to about -20.

Mechanisms of NTE based on transverse thermal motion of oxygens in two-fold coordination (W-O-Zr) have
been widely considered\cite{sle98,mary96,evans96,pryde96,ram98,ern98,david99},  however,
Cao et al.\cite{cao02,cao03}
have proposed an alternative model, based on XAFS data,
in which translation of \wo\ tetrahedra along the high-symmetry \direction{111} axes plays a key role.
Ultimately the discussion of mechanism centers on the eigenvectors of the relevant
phonon modes.
Toward this end, we have calculated the eigenvector for each \textbf{k}=0 optical phonon mode
using a mass-spring model of a 44 atom unit cell with nearest-neighbor stretching and bending interactions
and periodic boundary conditions. 
With appropriate bond strengths, the eigenvectors exhibiting bond stretching
and bending motion are associated with high (700-900\s\wn ) and moderate (150-400\s\wn ) frequency eigenvalues respectively, as expected.  For the range below about 120\s\wn\ our calculation generates 27 additional optic modes;
of which 21 are associated with triplets.  
The two lowest energy modes are triplets at 32 and 43\s\wn\  respectively, in reasonable correspondence
with the data.
The eigenvectors for these modes exhibit a mixture of librational and translational motion, as illustrated in Figure 3a.  
A suitable choice of eigenvector within the triplet degeracy manifold is required to clarify the motion
in this way. For the mode shown, clock-wise rotation about the \direction{111} axis 
is accompanied by downward translation of the two tetrahedra along the
\direction{111} axis\footnote{For most parameter choices, this is one of the two lowest energy modes.  
In the other the \direction{111}-axis octahedra rotates opposite to the two tetrahedra.
Two singlet modes with purely translational motion (along the \direction{111} axes) also occur at
relatively low frequencies which are sensitive to weak next-nearest-neighbor interactions.}  
  
This \direction{111} axis pair is connected, via Zr-centered octahedra, to other tetrahedra 
as illustrated in Figure 3b.  The are a total of eight \wo\ tetrahedra per unit cell
arranged in pairs along  \direction{111} axes and offset from each other in a ``spiral staircase'' pattern as shown.
Motion on the other \direction{111} is not the same\footnote{The motion includes librations  in which the unconstrained
oxygens tilt away from their \direction{111} axes.} 
and the motions are correlated in complex ways.

\begin{figure}
\begin{center}
\includegraphics[width=3.2in]{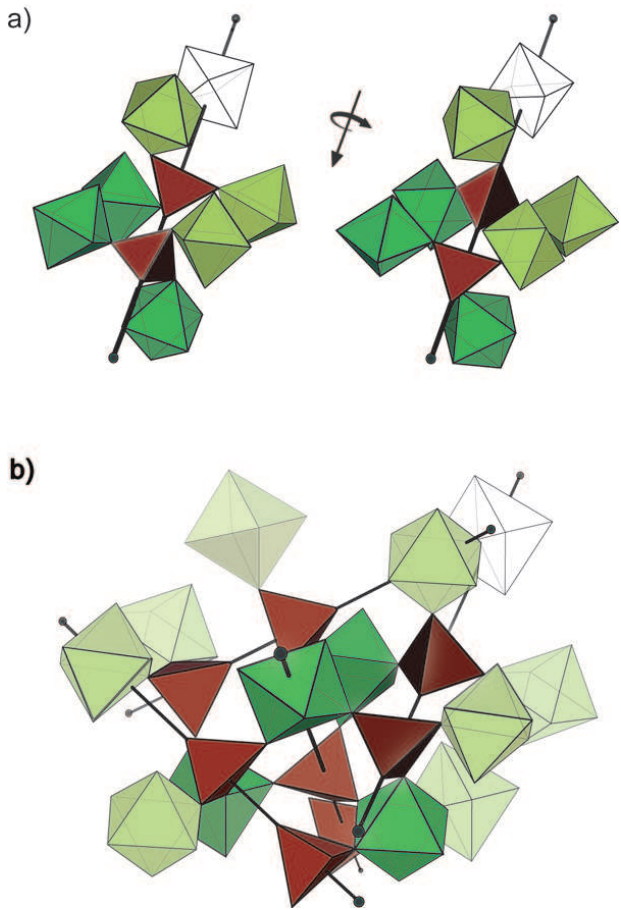}
\label{fig:ss}
\caption{(color online) a) The nature of a low frequency triplet mode is shown in terms of \wo\
tetrahedra and Zr-centered octahedra via a time sequence.
As the tetrahedra rotate clockwise, they translate downward
along the \direction{111} axis, which is shown by the dark line.    The two lowest-frequency optic mode are similar, but in the
other  one the \direction{111} axis
octahedron (white) rotates opposite the tetrahedra.  The unconstrained oxygen lies on the \direction{111}
axis.
b) The pair of tetrahedra shown in (a) couples to other tetrahedra as shown.  There are
eight \wo\ tetrahedra in a unit cell 
organized in pairs on \direction{111} axes (dark lines) which are offset in a spiral-staircase
configuration as shown. (Illustration by Alison Kendall)
}
\end{center}
\end{figure}

This mixing of translational and rotational (librational) motion is intimately 
related to the underconstrained nature of the \zwo\ structure, in which each tertrahedron
has an unconnected (terminal) oxygen at its
\direction{111} axis apex{\cite{ram98}. 
Because there are no strong second bonds for this unconstrained oxygen, 
translations of \wo\ tetrahedra along  \direction{111} axes do not involve significant bond compression 
and therefore project to very low frequency.  There they tend to mix with the low frequency librational motion, and it is
through this mixing that these low energy modes can acquire a dynamic dipole moment 
(since the \wo\ tetrahedra carry net charge).
Our observation of non-zero dipole moment for many low energy modes indicates
that this mixing is pervasive.
Thus the richness of the low frequency infrared spectra is connected
to the existence of the unconstained oxygen and the related openness 
along the high symmetry \direction{111} directions, which is a defining feature of the \zwo\ structure
and critical to NTE\cite{ram98,ern98,david99,mit99,allen03} .

Regarding the origins of anharmonicity in \zwo , we note that, in the context of
simple harmonic picture,  the uncertainties in the circumfrential oxygen position would be substantial.
Using the frequency of the lowest optic mode (28\s\wn) one gets a theoretical uncertainty in the ground
state of $\Delta\theta=\sqrt{2\hbar/m_{\tiny \textrm{O}} r^2 \omega}\simeq$14$\degree$ with a corresponding circumferential uncertainty $\Delta (r\theta)\simeq0.4\s\AA$.  These 
increase to about 40$\degree$ and $1.2\s\AA$ by room temperature.
Such large values suggest a possible origin of anharmonicity and motivate consideration of 
the relevance of configurational tunneling or
rotation to the low energy dynamics.
Tunneling, rotation and configurational exchange phenomenon have been observed in systems with a similar 3-fold rotation
axis involving
methyl-group tetrahedra\cite{prager97,dimeo02}, where the breakdown of the harmonic approximation for low-energy
librational motion has been carefully studied. For methyl-group tetrahedra such as \ch\  the lowest librational mode frequency
is typically 80\s\wn , about 3 times higher than  \zwo .  That most of this difference can be
accounted for by the hydrogen-oxygen mass difference suggests a roughly comparable confining
potential to lowest order.
A possible difference between \xch\ and \wo\ tetrahedra arises from
the nuclear spin, which is  $\frac{1}{2}$ for H, and $0$ for the O$^{16}$ nucleus, 
therefore a 120$\degree$ reorientation of an WO$_4$ tetrahedron is equivalent to a double permutation among 
identical particles and thus does not produce a distinct state.  
The potential relevance of 
finite barrier phenomena and large oxygen excursions to the the low-energy dynamics of
\zwo\  remains a question for future work.

We have provided evidence for  highly unusual low-energy phonon dynamics in \zwo\ as reflected 
in our infrared spectra.
We infer that the lowest optic modes in \zwo\ tend to have a mixed librational and translational character
in which the unconstrained oxygen plays an essential role, and
it is likely that these modes play a central role in the
mechanism of negative thermal expansion.  
While further work is needed to fully elucidate the mechanism of NTE in \zwo , 
the present works shows that both librational and translational motions are operative: the transverse oxygen motion 
associated with libration provides NTE, while translational component within each mode frustrates a displacive transition that would otherwise remove the unique structural environment in which these 
mixed phonon modes exist.

\begin{acknowledgments}
We gratefully acknowledge F. Bridges, D. Cao, O. Narayan and W.E. Pickett for helpful conversations.
Work at UCSC supported by NSF Grant DMR-0071949.
\end{acknowledgments}
\bibliography{zwo_refs1,jasonbooks}
\end{document}